\documentclass[10pt]{IEEEtran}
\usepackage{graphicx} 
\usepackage{amsmath}
\usepackage{braket}
\usepackage{textcomp}
\usepackage{cleveref}
\usepackage{fancyvrb}
\usepackage{amsfonts}
\usepackage{booktabs}
\usepackage{multirow}
\usepackage{float}
\usepackage{svg}
\usepackage{stfloats}
\usepackage{setspace}
\usepackage{pgfplots}
\usepackage{subcaption} 
\usepackage{amsmath}
\usepackage{tikz}
\usepackage{multirow}
\usepackage{multicol}
\usepackage{threeparttable} 
\usepackage[linesnumbered,ruled,vlined]{algorithm2e}

\usepackage{fancyhdr}
\newcommand*\circled[1]{\tikz[baseline=(char.base)]{
            \node[shape=circle,draw, inner sep=0.6pt, line width=1pt] (char) {#1};}}

\begin{document}

\title{ The Dilemma of Random Parameter Initialization and Barren Plateaus in Variational Quantum Algorithms}

\author{\IEEEauthorblockN{Muhammad Kashif \IEEEauthorrefmark{1}\IEEEauthorrefmark{2},
Muhammad Shafique\IEEEauthorrefmark{1}\IEEEauthorrefmark{2}}

\IEEEauthorblockA{\IEEEauthorrefmark{1}  eBrain Lab, Division of Engineering, New York University Abu Dhabi, PO Box 129188, Abu Dhabi, UAE}\\
\IEEEauthorblockA{\IEEEauthorrefmark{2} \normalsize Center for Quantum and Topological Systems, NYUAD Research
Institute, New York University Abu Dhabi, UAE}

Emails: \{muhammadkashif,muhammad.shafique\}@nyu.edu
    }

\maketitle
\thispagestyle{fancy}
\begin{abstract}
This paper presents an easy-to-implement approach to mitigate the challenges posed by barren plateaus (BPs) in randomly initialized parameterized quantum circuits (PQCs) within variational quantum algorithms (VQAs). 
Recent state-of-the-art research is flooded with a plethora of specialized strategies to overcome BPs, however, our rigorous analysis reveals that these challenging and resource heavy techniques to tackle BPs may not be required. Instead, a careful selection of distribution \emph{range} to initialize the parameters of PQCs can effectively address this issue without complex modifications. 
We systematically investigate how different ranges of randomly generated parameters influence the occurrence of BPs in VQAs, providing a straightforward yet effective strategy to significantly mitigate BPs and eventually improve the efficiency and feasibility of VQAs. 
This method simplifies the implementation process and considerably reduces the computational overhead associated with more complex initialization schemes.
Our comprehensive empirical validation demonstrates the viability of this approach, highlighting its potential to make VQAs more accessible and practical for a broader range of quantum computing applications. 
Additionally, our work provides a clear path forward for quantum algorithm developers seeking to mitigate BPs and unlock the full potential of VQAs.
\end{abstract}

\begin{spacing}{0.99}
\section{Introduction}
Quantum computing represents a significant leap from classical computing, harnessing the peculiar properties of quantum mechanics to process information in fundamentally different ways \cite{zaman2023survey}.
In recent years, significant advancements have been made in the field of quantum computing, resulting in the development and availability of noisy intermediate-scale quantum (NISQ) devices. Characterized by their relatively modest qubit counts, ranging from approximately $50$ to a few hundred, these devices are susceptible to errors stemming from quantum noise and decoherence \cite{Preskill_2018}. 
Despite these challenges, the emergence of NISQ technology has inspired substantial progress in the creation of post-quantum algorithms and applications specifically designed to operate effectively within the constraints of the NISQ regime \cite{kashif:2022a}.

Variational Quantum Algorithms (VQAs) are becoming a crucial framework for demonstrating quantum advantage in practical applications during the NISQ era, primarily due to their relatively low requirements for gate noise and circuit connectivity \cite{cerezo:2021,kashif:2021}.
VQAs operate within a hybrid quantum-classical framework utilizing parameterized quantum circuits (PQCs) \cite{Benedetti_2019}. 
In this setup, the quantum component incorporates a parameterized operation $V$ to encode the input data $x$ into a quantum state. 
This is followed by another parameterized operation $U(\theta)$, controlled by trainable parameters 
$\theta$, which modifies the encoded quantum states. Measurements of these states are subsequently performed to compute a cost function $C$.
The measurements yield classical results, and hence many state-of-the-art VQAs depend on the optimization of a PQC via a classical optimization loop \cite{McClean:2018}. 
Similar to the classical neural networks, VQAs typically use optimization methods such as gradient descent \cite{du:2019} and/or its variants \cite{bottou:2010} to iteratively adjust the parameters $\theta$. 
VQAs have been explored across various domains, including quantum simulations \cite{tabares:2023, robin:2023}, quantum chemistry \cite{lim:2024,xu:2024}, numerical analysis \cite{jaksch:2023,liu:2024}, and machine learning \cite{senokosov:2023,gong:2024}, demonstrating their versatility and broad impact in advancing quantum computing technologies.

Despite their widespread application,VQAs also face significant challenges \cite{kashif:2023_param_init_classical}. As the scale and complexity of the quantum circuits increases, in terms of qubit count and/or circuit depth, VQAs experience a trainability issue known as the barren plateaus (BP) problem \cite{McClean:2018,kashif:2024HQNET}. 
This issue is characterized by an exponential suppression of gradients across the parameter space in \emph{randomly} initialized VQAs that approximate to unitary 2-design\cite{Harrow_2009}, which significantly hinders the efficiency of gradient-based optimization methods in larger, more expressive, quantum systems \cite{Kashif:2023,kashif_unified}. 
Moreover, it has been demonstrated that even gradient-free optimization methods are not immune to the detrimental effects of BPs\ \cite{arrasmith:2021}, suggesting that the problem is inherent to the scaling of quantum circuits rather than specific to the type of optimization algorithm employed. Addressing the issue of BPs is therefore critical for the practical success of a wide range of applications utilizing VQAs.

\begin{figure*}
    \centering
    \includegraphics[scale=0.35]{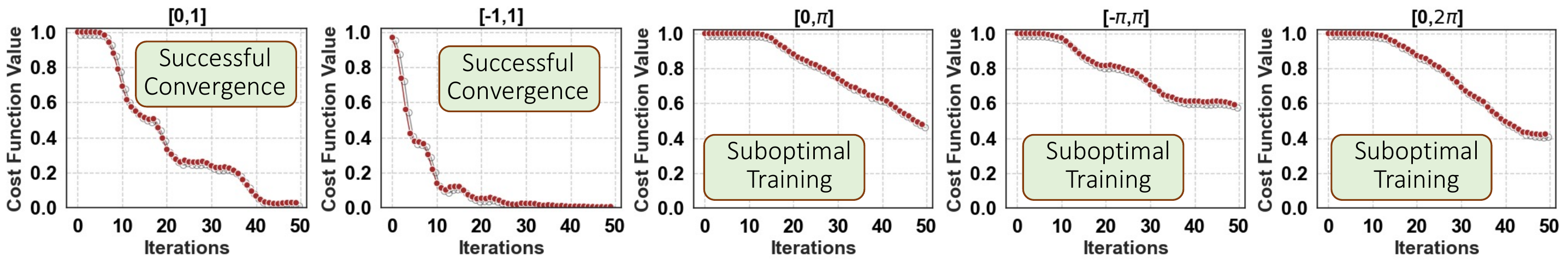}
    \caption{\footnotesize Impact of distribution range on the training performance of VQAs}
    \label{fig:motiv_analysis}
\end{figure*}
\subsection{Related Work}
A number of solutions have been proposed to alleviate the BP problem VQAs. Below we highlight some recently proposed solutions to address BPs and highlight their potential limitations:

\begin{itemize}

    \item \textbf{\textit{Use of State Efficient Ansatz:}} The state efficient ansatz (SEA), proposed in \cite{Xia:2024}, aims to mitigate BPs in VQAs by reducing parameter counts and enhancing trainability. However, its scalability to larger quantum systems remains uncertain, and it may encounter training challenges with an increasing number of qubits. Additionally, while SEA is effective for tasks like ground state estimation, its performance across other computational tasks can vary. The implementation of SEA also involves complex circuit designs to optimize entangling capabilities, which may be difficult to execute efficiently on NISQ devices due to their limited coherence times and high error rates.
    
    \item \textbf{\textit{Transfer Learning Inspired Parameter Initialization:}} A parameter initialization method inspired by transfer learning is proposed in \cite{Liu_2023} to mitigate BPs in training VQAs. However, the success of this method might heavily depend on the similarity between the initial small-sized task and the larger tasks to which the parameters are transferred. Secondly, the process of solving smaller tasks and then transferring parameters to larger ones might entail significant computational overhead, especially when dealing with a large number of preliminary tasks or complex parameter spaces.

    \item \textbf{\textit{Use of Guassian Mixture Model 
    for parameter Initialization:}}The use of a Gaussian Mixture Model for parameter initialization in VQAs, proposed in \cite{zhang:2022}, aims to mitigate BPs. While this method helps in escaping BPs, its effectiveness is largely dependent on the type of ansatz used, potentially limiting its applicability across different quantum circuit architectures. Additionally, implementing the Gaussian Mixture Model introduces computational complexity and overhead, requiring careful configuration, training of the model, and substantial expertise in both quantum computing and statistical modeling.

    \item \textbf{\textit{Residual Learning:}}A recent study suggested incorporating the residual approach to address the BP in quantum neural networks (QNNs) by dividing  in QNNs into multiple quantum nodes \cite{kashif2024resqnets,kashif:2024resqunns}. However, by segmenting QNNs into multiple nodes, each with its own PQC, the approach significantly increases the quantum hardware resources required. This includes a higher number of qubits and potentially more complex connectivity between them, which may not be feasible with the limited qubit availability and connectivity constraints of current NISQ devices.
 
\end{itemize}

The state-of-the-art solutions for addressing BPs in VQAs often present considerable challenges in terms of implementation and computational complexity.

\vspace{-10pt}
\subsection{Motivational Analysis}

In this study, we argue that constraining the distribution range for the initial parameters of PQCs in VQAs can mitigate the adverse effects of BPs. We illustrate this through a case study in which VQAs are trained to learn the identity function using a $15$-qubit PQC. We use the PQC architecture, characterized by an alternating layered structure initially proposed in \cite{Cerezo:2021aa}, which is particularly suitable for studying BPs. The same structure is used in the rest of paper also. 

Our findings reveal that minor adjustments to the distribution range of the initial parameters significantly influence the performance of VQAs for the same computational task, as depicted in Fig. \ref{fig:motiv_analysis}. \textit{This observation underscores the fact that the susceptibility of PQCs to BPs is not merely a consequence of random initialization but is strongly dependent on the chosen parameter range. These results highlight the necessity of further investigating which parameter ranges can effectively reduce the likelihood of encountering BPs, even with random initialization.}

\vspace{-10pt}
\subsection{Our Contribution}
In this paper, we introduce an alternative approach to alleviate adverse effects of barren plateaus (BPs) in randomly initialized parameterized quantum circuits (PQCs), a fundamental component in variational quantum algorithms (VQAs). 
We demonstrate through rigorous analysis that the commonly perceived necessity for specialized strategies to overcome BPs is not required. 
Instead, the issue can be effectively addressed by carefully selecting/restricting the range of \emph{randomly} generated parameters to initialize the gate parameters in PQCs being used in VQAs. 
We provide a systematic exploration of how different parameter ranges influence the likelihood of encountering BPs, thereby offering a straightforward yet effective method to enhance the efficiency and feasibility of VQAs. 
This method not only simplifies the implementation process but also significantly reduces the computational overhead associated with more complex initialization schemes. 
Through a comprehensive empirical validation, we underscore the viability of our approach, which promises to make VQAs more accessible and practical for broader applications in quantum computing.

\textbf{\textit{Summary of key results:}} Our results reveal that selecting initial parameters from a smaller distribution range can help mitigate the issue of barren plateaus (BPs). This is because restricting the parameter range prevents exploration of unnecessary regions in the solution space, reducing the likelihood of the optimizer becoming trapped in undesired local minima.
\vspace{-10pt}
\subsection{Organization}
The rest of the paper is organized as: 
Section \ref{sec:VQAs} presents an in-depth details of the working of VQAs. Section \ref{sec:BPs} explores the occurrence of BPs within VQAs. The methodology employed is detailed in Section \ref{sec:methodology}. Subsequently, Section \ref{sec:results} discusses the findings. Finally, Section \ref{sec:conclusion} provides conclusions drawn from the study.

\section{Variational Quantum Algorithms} \label{sec:VQAs}
To understand the BPs, we first  have to undertand the working of VQAs. A typical illustration of VQA design is presented in Fig. \ref{fig:vqa}. Below, we provide step-by-step details of VQAs:
\begin{figure}
    \centering
    \includegraphics[scale=0.42]{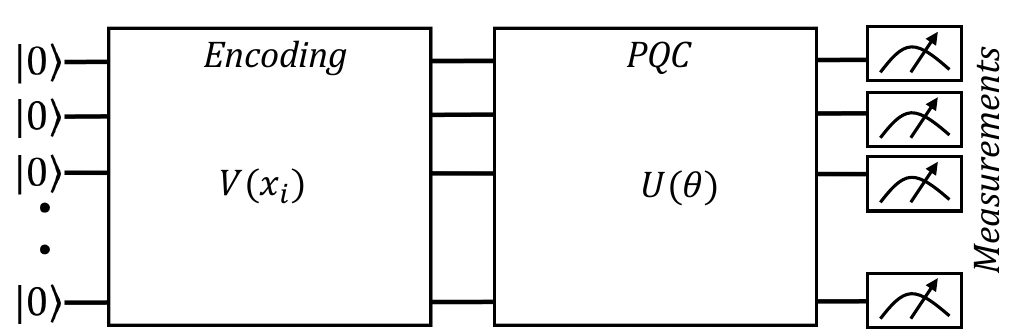}
    \caption{\footnotesize Typical VQA Architecture. In this exapmple the qubits are prepared in $\ket{0}$ state. VQAs have two sets of unitary transformations. The first unitary, $V$ , is used to encode the data into a quantum state. The second unitary, $U(\theta)$, is an arbitrary parameterization with parameters $\theta$, which are iteratively optimized during training to minimize the cost function $C$.}
    \label{fig:vqa}
\end{figure}

\vspace{-10pt}
\subsection*{1. Encoding}

The classical data  $x$  is encoded into a quantum state using a PQC $V(\phi, x)$:
\begin{equation}
    \ket{\psi(\phi, x)} = V(\phi, x) \ket{0}^n
\end{equation}
where $\ket{0}^n$ represents the initial state of $n$ qubits. The parameterization $V$ can adopt various forms, depending upon the encoding strategy employed, such as angle encoding or amplitude encoding. These methods translate classical data into distinct quantum mechanical representations: angle encoding converts data into qubit rotation angles, while amplitude encoding maps data to the qubit probability amplitudes. The details on data encoding is out of the scope of this work, however, interested readers can find more details on various data encoding methods in \cite{LaRose:2020,Schuld:2021a}.

\vspace{-10pt}
\subsection*{2. PQC}

Once the data is encoded, a PQC $U(\theta)$ is applied to the encoded state:
\begin{equation} \label{eq:PQC}
     \ket{\psi(\theta)} = \prod_{l=1}^L W_iU(\theta_i) \ket{\psi(\phi, x)}
\end{equation}
where
$$U(\theta) = \bigotimes_{j=1}^{n}R_\sigma(\theta_i^j)$$

where $i$ and $j$ represent layer and qubit respectively.  $R_\sigma(\theta)=e^{-i\theta_i^j\sigma/2}$ with $\sigma \in (\sigma_x, \sigma_y, \sigma_z)$ and $\theta$ are the parameters that will be optimized. $W_i$ are unparameterized gates and are generally multi-qubit gates primarily used for entangling the qubits which in general enhances the expressivity of the quantum circuit. 

\vspace{-10pt}

\subsection*{3. Measurements}

Measurements are made on the quantum state $\ket{\psi(\theta)}$ to obtain information used to compute the cost function:
\begin{equation}
    \text{Measurements} \rightarrow \text{Data for } C(\theta)
\end{equation}

\vspace{-10pt}

\subsection*{4. Cost Function}
The cost function is important in VQAs as it governs the optimization of parameters $\theta$. The specific choice of cost function usually depends on the task in hand, however, the cost function $C(\theta)$ is typically the expectation value of an observable $\mathcal{O}$:

\begin{equation}\label{eq:CF_VQA}
    C(\theta) = \langle \psi(\theta) | \mathcal{O} | \psi(\theta) \rangle
\end{equation}

\vspace{-30pt}

\subsection*{5. Optimization}

The parameters $\theta$ are iteratively optimized using a classical optimization algorithm to minimize the cost function $C(\theta)$:
\begin{equation}
    \theta_{\text{new}} = \theta_{\text{old}} - \eta \nabla_\theta C(\theta)
\end{equation}
where $\eta$ is the learning rate.


\section{Barren Plateaus} \label{sec:BPs}

Suppose that we have a PQC of the form as in Eq. \ref{eq:PQC} and the cost function defined as in Eq. \ref{eq:CF_VQA}. 
The gradient of $C(\theta)$ with respect to a parameter $\theta_k$ is computed as follows:
\[
\frac{\partial C}{\partial \theta_k} = \text{Re} \left( \langle \psi(\theta) | \mathcal{O} U^\dagger(\theta) \frac{\partial U(\theta)}{\partial \theta_k} | 0 \rangle \right)
\]
where $\frac{\partial U(\theta)}{\partial \theta_k}$ is the derivative of the unitary with respect to the parameter $\theta_k$.
If the circuit distribution from Eq. \ref{eq:PQC} forms a unitary 2-design \cite{Harrow_2009}, then the gradients of the cost function $C(\theta)$ tend to zero exponentially with the increase in the number of qubits or the depth of the circuit. Specifically:
$$\mathbb{E}\left[\left(\frac{\partial C}{\partial \theta_k}\right)^2\right] \propto \exp(-\alpha n)$$
for some positive $\alpha$, where $\mathbb{E}$ denotes the expectation over some distribution of the parameters $\theta$. The mathematical proof of this phenomenon is presented in \cite{McClean:2018,Friedrich:2022}.
However, a rather intuitive reasoning for the occurrence of BPs can be attributed to the increasing randomness in the quantum state space as the circuit becomes larger and more complex. Essentially, as the number of qubits and layers in a PQC increases, the resulting state $|\psi(\theta)\rangle$ approaches a state that is nearly orthogonal to minor variations induced by small changes in the parameters. 

When a state $\ket{\psi(\theta)}$ is nearly orthogonal to its variations $\ket{\psi(\theta+d\theta)}$ caused by small changes, it means the inner product $\bra{\psi(\theta)\ket{\theta+d\theta}}$ is close to zero. This near-zero inner product indicates that the state changes to one that is almost completely different from the original with even minor adjustments to the parameters.

In gradient-based optimization, which is often used to train PQCs, the gradient tells us how to change the parameters to achieve the desired optimization. If the state after a minor parameter change is nearly orthogonal to the original state, it means the sensitivity of the output state to parameter changes is very low. Consequently, the gradient (which depends on these sensitivities) becomes very small.
This behavior is a key feature of BPs in VQAs. If every small change leads to a state nearly orthogonal to the original, it means that the gradients are typically very small (close to zero), making it hard to find a direction in which to adjust the parameters to improve the performance of the quantum circuit. This drastically reduces the effectiveness of gradient-based optimization methods.



\section{Our Methodology} \label{sec:methodology}
We investigate how the range of initial parameters used to draw parameters from some random uniform distribution to set gate parameters of PQCs in VQAs affects the likelihood of encountering the BP problem. We experimented with different ranges, as shown in Table \ref{tab:param_ranges}.
According to the definition of BPs discussed in Section \ref{sec:BPs}, we use PQCs with substantial depth that approximate to a 2-design  while incrementally increasing the number of qubits. 
Our methodology is two-fold:
First, we analyze the variance of gradients, a key indicator of occurrence of BPs, with parameters of PQCs generated across different ranges. 
Secondly, we train the PQCs for a specific problem to assess potential learning-related benefits.
The experiments are carried out using PennyLane, a Python-based library designed for differentiable quantum computing. Below, we provide a detailed discussion of our methodology for both variance and training analysis.

\begin{table}[H]
\centering
\caption{\footnotesize Distribution Ranges Used to Initialize the Parameters of PQCs}
\label{tab:param_ranges}
\begin{tabular}{|l|l|} 
\hline
Parameters & Range     \\ \hline
Range 1   & \hfil $[0,1]$    \\ \hline
Range 2   & \hfil $[-1,1]$    \\ \hline
Range 3   & \hfil $[0, \pi]$  \\ \hline
Range 4   & \hfil $[-\pi,\pi]$   \\ \hline
Range 5   & \hfil $[0,2\pi]$  \\ \hline
\end{tabular}
\end{table}


\subsection{Method For Gradient's Variance Analysis}
The pseudocode provided in Algorithm \ref{algo:variance} outlines the procedure for computing the variance of gradients in PQCs. Below, we present a more detailed explanation of how this pseudocode operates:

\SetKwComment{Comment}{/* }{ */}
\begin{algorithm}[h]
\footnotesize 
\caption{Pseudocode to Compute Variance of 200 Random PQCs for different number of qubits implemented in Pennylane} \label{algo:variance}
\DontPrintSemicolon 
num\_samples $\gets$ $200$\;
qubits $\gets$ [$5,7,9,11,13,15$]\;
variances $\gets$ []\;
\SetKwFunction{FMain}{RandomCircuit}
\SetKwProg{Fn}{Function}{:}{}
\Fn{\FMain{params, random\_gate\_sequence, num\_qubits}}{
    \For{$i = 0$ \KwTo $num\_qubits$}{
        random\_gate\_sequence[$i$](params[$i$], wires=$i$)\;
    }
    \For{$i = 0$ \KwTo $num\_qubits - 1$}{
        qml.CZ(wires=[$i, i + 1$])\;
    }
    observable = qml.PauliZ(0)
    \For{$m=1$ \KwTo num\_qubits}{
    observable = observable @ qml.PauliZ($i$)
    }
    \KwRet qml.expval(observable)\;
}

\For{num\_qubits in qubits}{
    grad\_vals $\gets$ []\;
    random\_gate\_sequence $\gets$ \{\}\;
    \For{$i = 1$ \KwTo num\_samples}{
        dev $\gets$ qml.device("lightning.qubit", wires=num\_qubits)\;
        qcircuit $\gets$ qml.QNode(\FMain, dev, interface="autograd")\;
        grad $\gets$ qml.jacobian(qcircuit, argnum=0)\;
        gate\_set $\gets$ [$qml.RX$, $qml.RY$, $qml.RZ$]\;

        \For{$i$ \KwTo num\_qubits}{
        random\_gate = np.random.choice(gate\_set)
        random\_gate\_sequence[$i$] = random\_gate 
        }
        
        params $\gets$ np.random.uniform($0$, $1$, size=num\_qubits)\; 
        gradient $\gets$ grad(params, random\_gate\_sequence, num\_qubits)\;
        grad\_vals.append(gradient[-1])\; 
    }
    variances.append(np.var(grad\_vals))\;
}

variances $\gets$ np.array(variances)\;
\textbf{print}(variances)\;
\end{algorithm}

\subsection*{PQC Design for Variance Analysis}
The design of PQCs used for variance analysis (Line $4-11$ in Algorithm \ref{algo:variance}), consists of a layer of parameterized single-qubit gates followed by a ladder of controlled-Z ($CZ$) gates. The single-qubit gates are chosen randomly from the set {$RX$, $RY$, $RZ$}, which are Pauli-$X$, $Y$, and $Z$ rotation gates, respectively. 
Each gate is parameterized by a single initial parameter, sampled uniformly from different ranges, as shown in Table \ref{tab:param_ranges}. The connectivity of the $CZ$ gates follows a nearest-neighbor pattern, wherein each qubit is entangled with its subsequent qubit except for the last qubit.
The observable measured in each circuit is the tensor product of Pauli-$Z$ operators applied across all qubits. The expectation value of the observable is computed, providing the output of the circuit.
The PQC returns a flattened array representing the probabilities of measuring each computational basis state 
 \(|i\rangle\) given the quantum state \(|\psi\rangle\), i.e., \(|\langle i|\psi\rangle|^2\). The PQC is repeated twice before until the measurement in all the experiments.


\subsection*{Variance Computation}

To evaluate the variance in gradients, we generate $200$ random PQCs, for each specified number of qubits (Line $12-25$ in Algorithm \ref{algo:variance}). The initial parameters for these gates are drawn from a uniform distribution, with  systematically varying the range of distribution as summarized in Table \ref{tab:param_ranges}. 
For each individual PQC, a unique random sequence of single-qubit gates is created, and the circuit is executed. The gradient of each PQC's expectation value with respect to the parameters of the single-qubit gates is the computed (Line $18$ in Algorithm \ref{algo:variance}). 
We then extract the final component of the gradient vector from each run and compute the variance (Line $24$ in Algorithm \ref{algo:variance}).

The experiments are performed across various PQC sizes with $5, 7, 9, 11, 13$, and $15$ qubits to determine how system size influences the variance of gradient estimates, across different ranges from which the initial parameters are drawn.
The calculated variances from these simulations are stored and subjected to further analysis to identify emerging trends and patterns. This analysis aims to provide insights into the scalability and stability of gradient computations which is an important indicator of BPs in quantum circuits, and is essential for optimizing quantum algorithms.


\vspace{-10pt}
\subsection{Method For Training Analysis}

We then analyze the training dynamics of PQCs by training them to learn the identity operation for different numbers of qubits and various parameter initialization ranges. The pseudocode for training analysis is outlined in Algorithm \ref{algo:training}.

\SetKwComment{Comment}{/* }{ */}
\begin{algorithm}[h]
\footnotesize
\caption{Pseudocode for training analysis with different parameter initializations in pennylane. The cost function used is to learn the identity gate function }
\label{algo:training}
\DontPrintSemicolon

\SetKwInput{KwInput}{Input}
\SetKwInput{KwOutput}{Output}
\KwInput{wires = $x$, layers = $2$, steps = $50$, num\_runs = $10$} \Comment{the wires denotes the number of qubits and can be set accordingly in each experiment}
\KwOutput{mean\_costs}

\SetKwFunction{FQuantumNode}{PQC}
\SetKwProg{Fn}{Function}{:}{}
\Fn{\FQuantumNode{weights}}{
    \For{$i = 0$ \KwTo $wires$}{
        qml.RY(weights[i][0], wires=i)\; 
    }
    \For{$l = 0$ \KwTo $layers$}{
        \For{$d = 0$ \KwTo $wires - 1$ }{
            qml.CZ([d, d + 1])\;
        }
        \For{$w = 0$ \KwTo $wires - 1$}{
            qml.RY(weights[w][1], wires=w)\;
        }
        \For{$x = 1$ \KwTo $wires - 2$ }{
            qml.CZ([x, x + 1])\;
        }
        \For{$w = 1$ \KwTo $wires$}{
            qml.RY(weights[w][2], wires=w)\;
        }
        observable $\gets$ qml.PauliZ(0)\;
        \For{$i = 0$ \KwTo $wires$}{
            qml.probs(i))\;
        }
    }
    \KwRet qml.expval(observable)\;
}
\Comment{Cost Function}
\SetKwFunction{FCost}{cost\_ftn}
\SetKwProg{Fn}{Function}{:}{}
\Fn{\FCost{weights}}{
    out $\gets$ $1 - (\FQuantumNode(weights))^2$\;
    \KwRet out\;
}

dev $\gets$ qml.device("$lightning.qubit$", wires=wires)\;
QNODE\_1 $\gets$ qml.QNode(\FQuantumNode, dev)\;

params $\gets$ np.random.uniform(0, 1, size=(wires, layers))\;  \Comment{Initial parameters in this example are generated in range 0 to 1 and can be changed accordingly to other ranges used in this paper}
params $\gets$ np.array(params, requires\_grad=True)\;
opt $\gets$ qml.AdamOptimizer(stepsize=0.1)\;
average\_costs $\gets$ []\;
\Comment{Training Loop}
\For{run = 0 \KwTo num\_runs-1}{
    print("Experiment", run)\;
    costs $\gets$ []\;
    \For{i = 0 \KwTo steps }{
        params $\gets$ opt.step(\FCost, params)\;
        print("Cost after step", i + 1, ":", \FCost(params))\;
        costs.append(\FCost(params))\;
    }
    average\_costs.append(costs)\;
}

mean\_costs $\gets$ np.mean(average\_costs)\;

\end{algorithm}

\subsection*{PQC Design for Training Analysis}
\begin{figure}[h]
    \centering
    \includegraphics[scale=0.27]{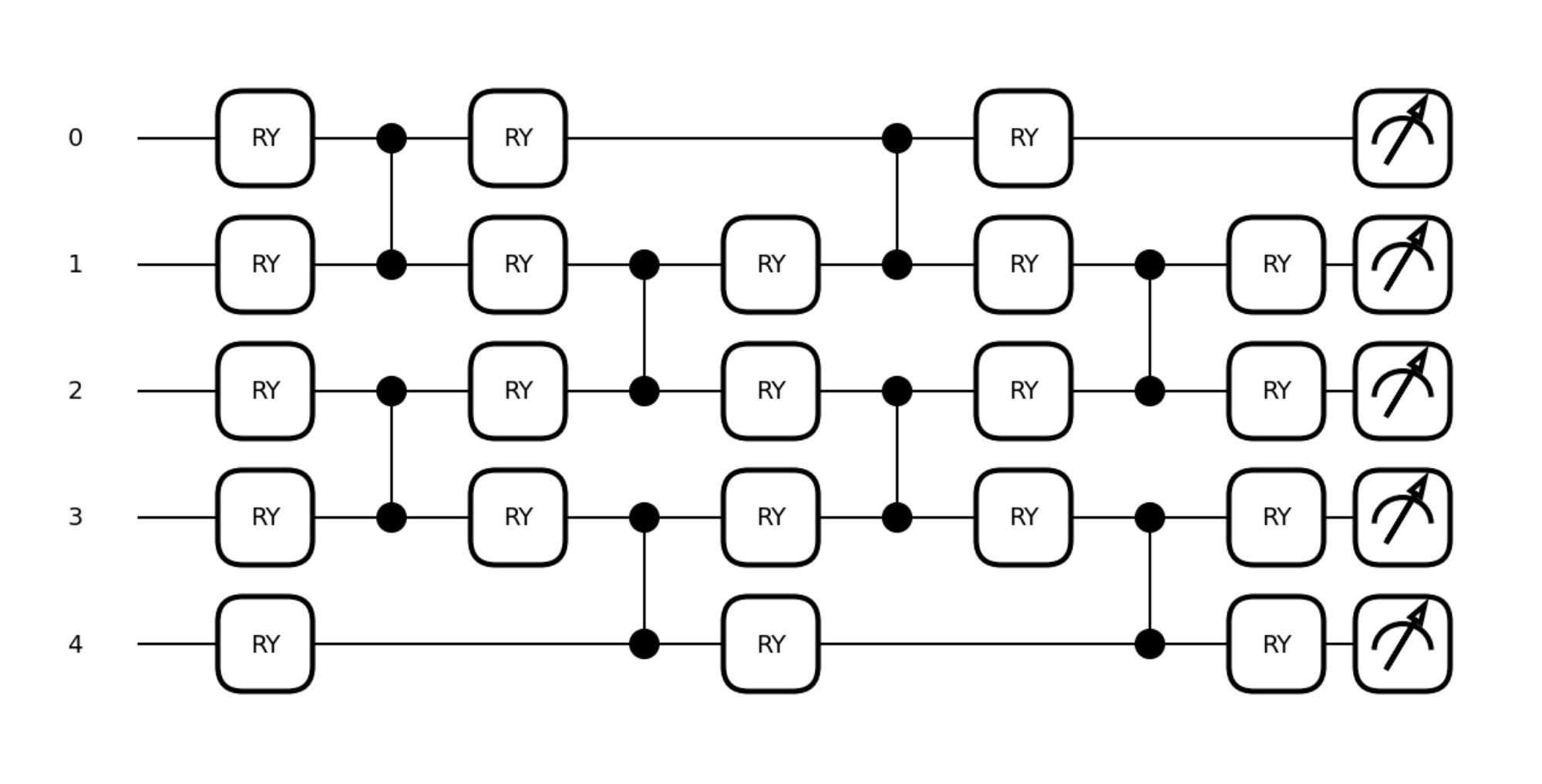}
    \vspace{-22pt}
    \caption{\footnotesize The Design of Parameterized Quantum Circuit used for Training Analysis. Inspired from \cite{Cerezo:2021aa}}
    \label{fig:PQC}
\end{figure}

For the training analysis, we used the PQC design proposed in \cite{Cerezo:2021aa}, schematic of which is shown in Fig. \ref{fig:PQC}, and the corresponding pseudocode is  from Line $1-16$ in Algorithm \ref{algo:training}. 
This particular PQC design does not strictly form a $2$-design, as it does not consist of universal $2$-qubit gates as building blocks, but has been shown to exhibit important properties to study BP in VQAs.

\subsection*{Cost Function}
The objective of the training is to minimize the deviation of the circuit's output from the identity operation (Line $17-19$ in Algorithm \ref{algo:training}). The cost function is defined as $1 - PQC(weights)^2$, where 
$PQC(weights)$ is the expectation value of the observable from the PQC. This cost function effectively measures that given a quantum state initialized by the input parameters, how close the optimization loop brings it to the identity operation.


\begin{figure*}[h]
    \centering
    \includegraphics[scale=0.5]{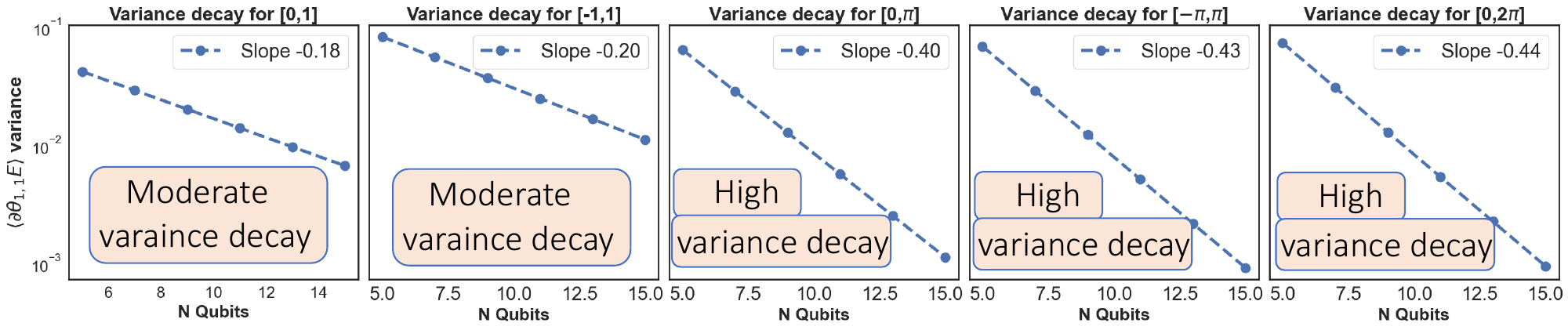}
    \vspace{-10pt}
    \caption{\footnotesize Variance of VQAs with parameters initialized randomly from uniform distribution in different ranges. Moderate variance decay denotes the setting which is less likely to encounter the BPs, whereas higher decay denotes the setting which is more susceptible to encounter BPs.}
    \label{fig:variance_fig_Probs}
    \label{fig:}
\end{figure*}

\subsection*{Training with Statistical Noise Consideration}
The training uses the Adam optimizer with a step size of $0.1$, and the initial parameters of parameterized gates in PQCs are generated using a random uniform distribution in different ranges (Table \ref{tab:param_ranges}), to explore how different initial conditions impact learning efficacy and convergence behaviors.
It is important to mention that the expectation value (observable) is estimated using finite shots or samples. Due to this finite number of samples, there is inherent statistical noise in each expectation value calculation. This noise can cause fluctuations in the final outcome of PQC, which consequently can effect the optimization trajectory.
To address this, instead of focusing on individual experimental runs, we conduct experiments across $10$ independent runs to ensure statistical validity, with each run comprising $50$ training steps. The training results are then averaged to get a better understanding of typical behavior.


\section{Results and Discussion} \label{sec:results}

\subsection{Variance Results} \label{sec:var_results}
The results depicted in the Fig. \ref{fig:variance_fig_Probs} highlight the variance decay in PQCs (a fundamental component of VQAs), as a function of the number of qubits. These results were obtained by varying the range of randomly initialized parameters in different PQCs in Algorithm \ref{algo:variance}.
Variance decay is a key indicator of BPs, where higher variance decay suggests the presence of such plateaus, leading to gradient-based training becoming inefficient.
Each subplot in the figure shows the following:

\subsection*{Parameter Ranges and Variance Decay}
\textbf{Range [$0,1$] and [$-1, 1$]:}  
When the parameters of PQCs are initialized randomly from relatively narrower distributions, specifically, from $0$ to $1$ and from $-1$ to $1$, the decay in the variance of the gradients exhibits slopes of approximately $-0.18$ and $-0.20$, respectively. These values suggest a moderate rate of variance decline. The slower decrease in variance implies a significantly reduced likelihood of encountering the BP problem in VQAs utilizing these initialization ranges. Consequently, such initialization schemes provide greater potential for effective optimization of the PQCs, eventually enhancing the overall performance of VQAs.

\textbf{Range [$0, \pi$], [$-\pi, \pi$] and [$0, 2\pi$]:} 
When the parameters of PQCs are initialized from the range [$0, \pi$], the slope of the variance decay is steep, at approximately $-0.40$. This indicates a pronounced decline in variance, suggesting a higher likelihood of encountering barren plateaus, which can complicate optimization processes.

For the range [$-\pi, \pi$], the decay slope is roughly $-0.43$, showing a significant reduction in variance as the number of qubits increases. This trend points to increased challenges associated with barren plateaus, potentially impacting the efficiency of variational quantum algorithms (VQAs).

Lastly, the range [$0, 2\pi$] demonstrates the steepest slope of approximately $-0.44$. This suggests that this initialization range experiences the most rapid variance decay among the tested ranges, indicating the highest probability of facing barren plateaus and posing severe challenges for effective optimization in VQAs.




The results indicate that different parameter initialization ranges lead to varying degrees of variance decay in VQAs. In general, wider ranges (like [$0, \pi$], [$-\pi, \pi$] and [$-\pi, \pi$]) tend to have steeper slopes, suggesting a higher risk of BPs. Narrower ranges, such as [$0, 1$] and [$-1, 1$], show a more moderate decline in variance, indicating a potentially lower risk of encountering BPs.

These findings provide crucial insights into how the choice of parameter initialization range in PQCs can potentially influence the performance of VQAs by affecting the occurrence of BPs. We can achieve approximately $55-60\%$ improvement in variance just by restricting the range of randomly generated initial parameters. Initializing the PQC's parameters in narrow ranges would be beneficial, leading to enhanced stability and performance in VQAs, thereby reducing computational overhead and improving overall feasibility.

\vspace{-10pt}
\subsection{Training Analysis}
\begin{figure*}
    \centering
    \includegraphics[scale=0.35]{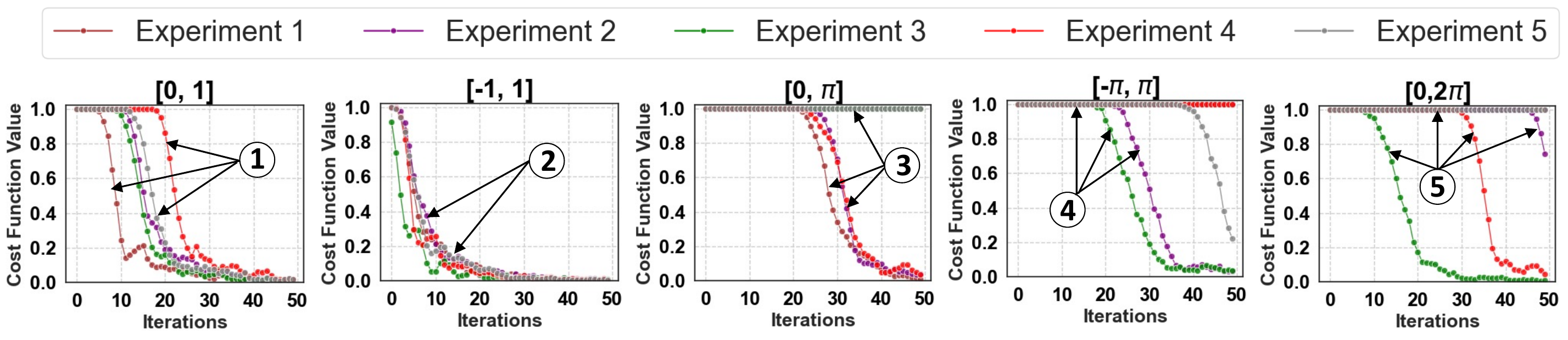}
    \caption{\footnotesize Impact of Statistical Noise on VQAs Training Performance. Different experimental iterations for same range of initial parameters yields significantly different results. To demonstrate this, we used $15$ qubit PQCs of the form as in Fig. \ref{fig:PQC}. The training is performed for $50$ iterations and Adam optimizer with learning rate of $0.1$ is used for cost function optimization.}
    \label{fig:statistical_noise}
\end{figure*}


\subsection*{The Need for Statistical Noise Consideration}
We observe that the statistical noise due to the finite number of measurement shots can significantly affect the training performance of VQAs. Specifically, the outcomes from different independent training iterations can vary considerably, indicating a high sensitivity to measurement noise. This variability is illustrated in Figure \ref{fig:statistical_noise}, where fluctuations in the training results demonstrate the inherent noise in quantum computations primarily due to measurement results fluctuations leading to the inconsistent convergence and performance in VQAs, suggesting a need for robust techniques to mitigate these effects.

To counter the challenges posed by statistical noise, we conducted $10$ independent training runs for each parameter initialization range considered in this paper. The results are then averaged to gain a clear perspective of the overall training performance of VQAs with parameters initialized across varying ranges. This helps to reduce the impact of measurement noise, providing more reliable insights into the behavior of VQAs during training.

Another notable observation from statistical noise consideration is that is that smaller parameter generation ranges, such as $[0,1]$ and $[1,1]$, not only significantly reduce variance decay (see Fig. \ref{fig:variance_fig_Probs}) bit also demonstrate a higher resistance to the statistical noise introduced by measurements in PQCs, as indicated by labels \circled{1} and \circled{2} in Fig. \ref{fig:statistical_noise}. In contrast, larger parameter ranges, such as, $[0, \pi]$, $[-\pi, \pi]$, and $[0, 2\pi]$ are more susceptible to statistical noise, often resulting in inconsistent performance across different experimental runs, as shown by labels \circled{3}, \circled{4}, and \circled{5} in Fig. \ref{fig:statistical_noise}.


\subsection*{Training Results of Randomly Initialized VQAs with Different Ranges}
\begin{figure*}
\centering
    \includegraphics[scale=0.35]{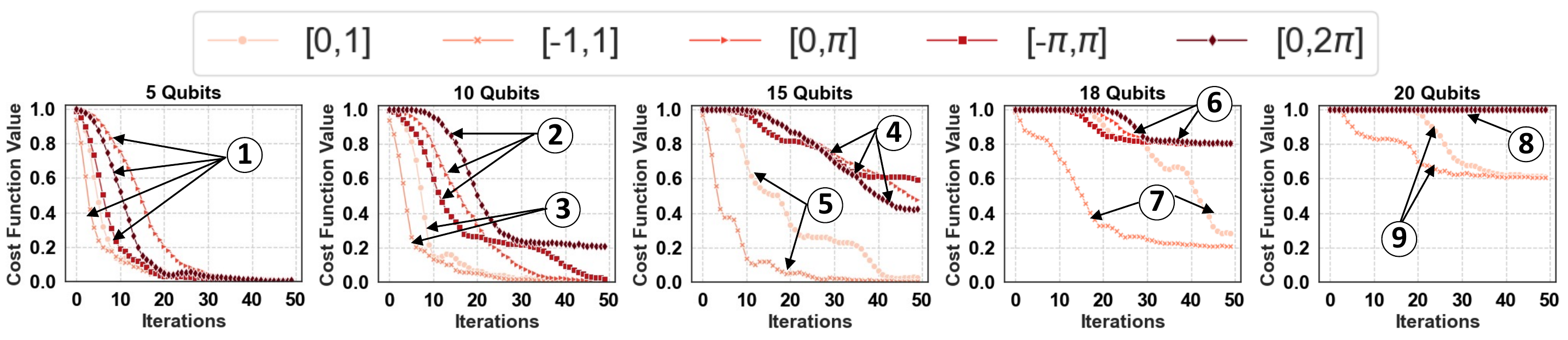}
    \caption{\footnotesize Training Results of VQAs. It can be observed that progressively increasing the number of qubits in the underlying PQCs while controlling the distribution range of initial parameters can mitigate BPs. Specifically, restricting the range from which parameters are drawn enhances the training potential of VQAs up to $20$ qubits. Conversely, larger initial parameter ranges tend to result in VQAs encountering BPs more rapidly, thus impeding effective optimization as the system size increases. }
    \label{fig:res_train}
\end{figure*}
We observed and demonstrated a significant improvement in variance decay by restricting the range of randomly initialized parameters in Section \ref{sec:var_results}. This observation is critical, as it suggests that limiting the parameter range may reduce the likelihood of encountering BPs in VQAs. 
However, it is also essential to consider whether these improvements can be translated in practical tasks, specifically when VQAs are trained to learn a particular task. To investigate this, we trained VQAs with various initial parameter ranges to learn the identity function. The results are presented in Figure \ref{fig:res_train}.
For training, we used PQCs with qubit counts of $5,10,15,20$ and $25$. The Adam optimizer was used for optimization, with a learning rate of $0.1$, across $50$ iterations. This allows us to assess the impact of parameter initialization ranges on the convergence and performance of VQAs in learning tasks.

For VQAs containing PQCs with a smaller qubit count, specifically in $5$-qubit PQCs, the initial parameter range appears to have minimal impact on training outcomes. Across all parameter ranges, these VQAs successfully trained without encountering the BP problem, as indicated by label \circled{1} in Fig. \ref{fig:res_train}. However, smaller initialization ranges, such as $[0, 1]$ and $[-1, 1]$, demonstrate slightly faster training times compared to larger ranges like $[0, \pi]$, $[-\pi, \pi]$, and $[0, 2\pi]$, typically reflecting their improved variance decay.

According to the definition of BPs, an increase in the number of qubits leads to a flattening of the cost function landscape which leads to trainability problems. In our experiments, when the qubit count is increased to 10, we observe that trainability issues starts to arise, particularly for larger initialization ranges ($[0, \pi]$, $[-\pi, \pi]$, and $[0, 2\pi]$). In these cases, training performance marginally deteriorates, as evidenced by slower and suboptimal convergence, highlighted by label \circled{2} in Fig. \ref{fig:res_train}. This decline in performance is typically associated with a more rapid decay in variance (as shown in Fig. \ref{fig:variance_fig_Probs}). Conversely, smaller initialization parameter ranges still enable successful training, as indicated by label \circled{2} in Fig. \ref{fig:res_train}.

Increasing the number of qubits to $15$ and $18$ leads to a substantial decline in performance when larger parameter initialization ranges are used. This decline results in suboptimal to negligible training outcomes, as indicated by label \circled{4} and \circled{6} in Fig. \ref{fig:res_train}, respectively. However, with smaller initialization ranges, successful training and convergence to a solution are still achieved, although with a slightly longer convergence time at higher qubit counts. This successful convergence, as highlighted by label \circled{5} and \circled{7} in Fig. \ref{fig:res_train}, demonstrates a strong resistance to BPs while using smaller initialization ranges even with a greater number of qubits.

Increasing the number of qubits to $20$ leads to no training at all when the initial parameters of VQAs are sampled from larger intervals such as $[0, \pi]$, $[-\pi, \pi]$, and $[0, 2\pi]$. This is illustrated in Figure \ref{fig:res_train}, labeled as \circled{8}. On the other hand, VQAs with parameters randomly initialized from smaller intervals, while yielding suboptimal results, still demonstrate some level of training, as indicated by label \circled{9} Fig.\ref{fig:res_train}, which shows the training potential and can be further enhanced with tunning of other hyperparameters.


\section{Conclusion}\label{sec:conclusion}
This paper presents an efficient approach to mitigate the impact of barren plateaus (BPs) in variational quantum algorithms (VQAs) by optimizing the initialization range of parameters in parameterized quantum circuits (PQCs). Our findings removes the need to complex and resource-intensive state-of-the-art techniques, necessary to overcome BPs. Instead, we demonstrate that a simple yet effective adjustment of the parameter initialization range can significantly reduce the likelihood of encountering BPs, improving the performance and feasibility of VQAs. 

Our systematic investigation into the influence of different parameter ranges on BPs shows that carefully restricting these ranges can significantly reduce the likelihood of BPs occurrence, thereby enhancing the stability and efficiency of VQAs without introducing additional computational complexity. This is because restricting the parameters range avoids the exploration of unnecessary solution that may lead the optimizer to get stuck unwanted local minima region.
%
The empirical validation of our approach highlights its effectiveness, underscoring its potential to simplify the implementation process and reduce computational overhead. This work contributes a practical solution to the BP problem and lays the foundation for further exploration and optimization of VQAs, offering a promising path forward for quantum algorithm development. Future research should explore the broader application of this method across various quantum circuits and problem domains, further unlocking the potential of VQAs in quantum computing.

\section*{Acknowledgements}
\vspace{-0.14cm}
This work was supported in part by the NYUAD Center for Quantum and
Topological Systems (CQTS), funded by Tamkeen under the NYUAD Research
Institute grant CG008.

\end{spacing}


    
\bibliographystyle{ieeetr}
\bibliography{main.bib}
\end{document}